# 2nd Workshop on Hybrid Development Approaches in Software Systems Development

Marco Kuhrmann[1(✉)], Philipp Diebold[2], Stephen MacDonell[3], and Jürgen Münch[4]

[1]Department of Computer Science, Institute for Applied Software Systems Engineering, Clausthal University of Technology, Goslar, Germany
marco.kuhrmann@tu-clausthal.de
[2]Fraunhofer IESE, Kaiserslautern, Germany
Philipp.Diebold@iese.fraunhofer.de
[3]Auckland University of Technology, Auckland, New Zealand
stephen.macdonell@aut.ac.nz
[4]Reutlingen University, Bögblingen, Germany
Juergen.Muench@Reutlingen-University.de

**Abstract**

*Software and system development is complex and diverse, and a multitude of development approaches is used and combined with each other to address the manifold challenges companies face today. To study the current state of the practice and to build a sound understanding about the utility of different development approaches and their application to modern software system development, in 2016, we launched the HELENA initiative. This paper introduces the 2nd HELENA workshop and provides an overview of the current project state. In the workshop, six teams present initial findings from their regions, impulse talk are given, and further steps of the HELENA roadmap are discussed.*

## 1. INTRODUCTION

Practitioners face numerous challenges in selecting the appropriate development approach for an organization, a team or a project. Since there is no "Silver Bullet" [2] in software system development, software engineers are on the quest for suitable development approaches, yet facing a huge variety of dynamic contextual factors influencing the definition of appropriate processes [3,10]. Hence, a variety of development approaches compete for the users' favor: standard approaches as well as home-grown approaches, more traditional and/or more agile ways of work, and projects influenced by the need to adhere to standards, norms, or regulations.

In 2015, West claimed that "Water-Scrum-Fall" had become reality [9]. A systematic review to investigate the current state-of-practice in software process use [8] revealed a considerable imbalance in the research concerning traditional and agile software system development. As a consequence, we initiated HELENA that aims to study the use of "Hybrid dEveLopmENt Approaches in software systems development". This initiative grew to a real project involving about 80 researchers[1] from (currently) 26 countries. Each of these 26 sites has a local head supporting the general organization team, and we owe special thanks to all our colleagues, who helped us quality assuring the survey instrument, translating the instrument, and spreading the word among their local peers. Initial results—in particular from the HELENA trials and the first stage of the study—have been presented at the annual meeting of the *Software Process* special interest group of the German Computer Society [6], at the *International Conference on Software System Process* (ICSSP) 2017 [4], and in [5].

The remainder of the paper is organized as follows: In Sect. 2, we provide a brief overview of the current state of the study. Section 3 introduces the workshop as such, and Sect. 4 provides a summary of future work.

## 2. THE HELENA STUDY: OVERVIEW AND CURRENT STATE

In this section, we provide a quick overview of the current state of the HELENA study from a global perspective. Information provided concerns the current dis- semination of the survey (Sect. 2.1) and few selected results (Sect. 2.2) grounded in a data dump from mid August 2017. Furthermore, detailed results can be obtained from the region-specific reports, which are introduced in Sect. 3.

---
[1] The full list of all HELENA contributors can be depicted from: https://helenastudy.wordpress.com/helena-team.

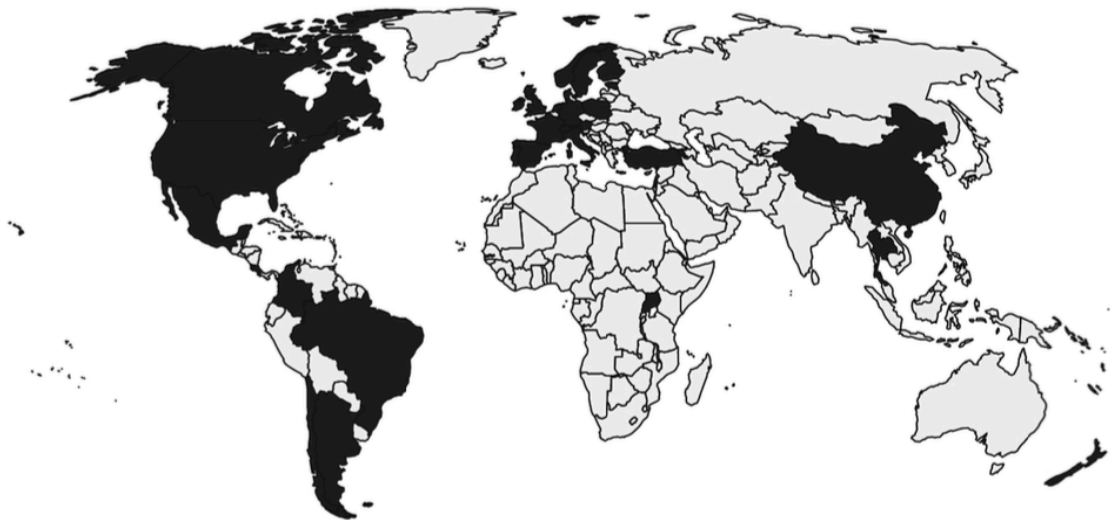

Fig. 1. Overview of the countries from which we received answers to the questionnaire (status: August 15, 2017).

## 2.1 Current State

The second HELENA workshop aims at discussing preliminary results from HELENA's stage 2. For this, the participating teams were invited to provide an initial analysis of a dataset, which was dumped from the survey tool on August 15, 2017. This dataset comprised 501 complete[2] data points, i.e., completely answered questionnaires. Figure 1 illustrates the countries from which we received answers. In total, we received data points from 31 countries. Among these data points, more than 20 data points each come from 10 countries and 14 countries provided fewer than 5 data points. The stage 2 questionnaire of HELENA was made available to the public on May 2, 2017 and accepts answers until September 30, 2017.

## 2.2 Selected Results

As already found in [4], the HELENA dataset is rich with information. Thus, in this section, we only provide a quick overview of selected results. The questionnaire was made available in English (35% of all answers), German (26%), Spanish (25%), and Portuguese (14%). Included in this analysis are the "complete" answers only, which results in an $n = 501$. Yet, since the questionnaire has a number of optional and multiple-selection questions, we have a varying $n$, which is reported in the respective answers. In the following, we provide some basic parameters:

- **Company Size** ($n = 501$): We provided five categories for the company size from which the participants could choose: micro-sized (11.58% of the participants), small (14.97%), medium (27.54%), large (23.95%), and very large (21.76%). Only 0.2% of the participants did not provide information regarding the company size.

- **Distributed work** (n = 501): The participants were asked to state their dis- tributed work pattern. In total 38.12% of the participants stated that they do not work in a distributed manner, 25.75% use distributed work within the same country, 11.98% in the same region, i.e., the same continent. Finally, 23.95% use globally distributed work. Again, 0.2% of the participants did not provide information.

- **Product/Project Size** ($n = 501$): A considerable share of the participants classifies the projects they refer to in their answers as very large, i.e., more than one person year in effort (60.88%). For the remaining categories, we received the following answers: large: 17.76%, medium: 15.37%, small (less than one person month): 4.19%, and very small (less than two person weeks): 1.8%.

- **Experience** ($n = 501$): In total, 63.07% of the participants stated that they have more than 10 years of experience. Another 15.97% have 6–10 years, 13.97% have 3–5 years, and 4.59% of the participants has 1–2 years of experiences. Only 2.40% mentions to have less than one year of experience.

In the survey, we asked the participants if they (intentionally) combine different development approaches, and 74.85% positively answered this question. In the regard, we are interested—similar to [4]—in the self-perception of the participants' way of work. To this end, we asked the participants to rate their way of implementing the standard SWEBoK disciplines [1]. Figure 2 shows that the participants aim at implementing a balanced process ecosystem, yet with a strong tendency toward agile.

Concerning the development approaches as such, we provided the participants with two lists: one comprising 24 (large, integrated) development methods and frameworks, and a second list comprising 35 techniques and practices. We did not provide an explicit categorization, whether a method or a practice is "traditional" or "agile", but provided the alphabetically sorted lists only. In total, we received 30,060 selections on the 7-point Likert scale answers that will help us to identify particular combination patterns. Just these few pieces of information show the richness of the HELENA dataset; further exciting insights are reported by the presenters of this workshop.

---

[2] It has to be mentioned that we have more that 1,100 data points in the database. Nevertheless, for the preliminary analyses, we only include those data points from completed questionnaires.

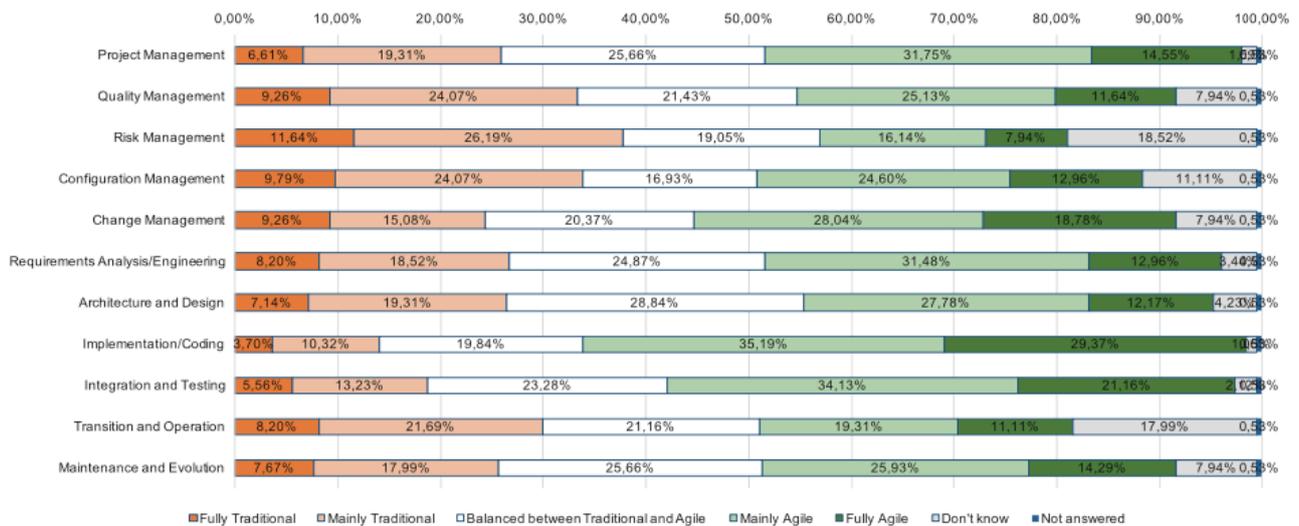
Fig. 2. Self-evaluation of the participants concerning the implementation of the standard SWEBoK disciplines ($n$ = 378).

## 3. The Workshop

This 2nd HELENA workshop aims at continuing the community work initiated at ICSSP 2016 (Austin, Texas); in particular, the HELENA survey. It continues the 1st workshop held in conjunction with ICSSP 2017 (Paris; [7]).

### 3.1 Overview

In this workshop, we aim at bringing together all academic and industry contributors and further interested people to:

1. Report the current state and (preliminary) outcomes of the HELENA survey
2. Develop a work program and define next steps within the whole community
3. Build working groups to work on selected (sub-)topics of interest
4. Create a research agenda for hybrid software and system development

This second workshop comprises reports from the regions presenting the current state of the data collection and analysis, posters that report status and/or present research questions, and (external) "lightning" talks given by researchers and practitioners not involved on the HELENA core activities to challenge the HELENA community. For the regions' reports, we asked the regions to submit short position papers, which were thoroughly reviewed by the HELENA core coordination group. Six regional and cross-regional papers have been invited for presentation. Hence, this second workshop also aims at informing the research community as well as practitioners about the current state of practice.

Table 1. Overview of the workshop topics and schedule.

| No. | Topic |
|---|---|
| 1 | *Introduction (Organizers)* |
| 2 | Report of the current state from a global perspective (Organizers) |
| 3 | Reports from the regions *(followed by a discussion of the global and regional status in the whole group)*: <br> 1. Initial Results of the HELENA Survey Conducted in Estonia with Comparison to Results from Sweden and Worldwide; *Ezequiel et al.* for Estonia and Sweden <br> 2. Hybrid Software and Systems Development in Practice: Perspectives from Sweden and Uganda; *Nakatumba-Nabende et al.* for Uganda and Sweden <br> 3. HELENA Stage 2—Danish Overview; *Tell et al.* for Denmark <br> 4. HELENA Study: Reasons for Combining Agile and Traditional Software Development Approaches in German Companies; *Klünder et al.* for Germany <br> 5. Hybrid Software and System Development in Practice: Initial Results from Austria; *Felderer et al.* for Austria <br> 6. HELENA Study: Initial observations of Software Development Practices in Argentina; *Paez et al.* for Argentina |
| 4 | Wrap up of results and work that happened since the first workshop |
| 5 | Lightning talks and poster presentations |
| 6 | Setup of working groups and continuing work in breakout session |
| 7 | Presentation of the working groups and their plans |
| 8 | Development of the HELENA Agenda and next steps |
| 9 | *Closing (Organizers)* |

## 3.2 Workshop Organization

The 2nd HELENA workshop is a 1-day workshop aiming at bringing together all the contributors of the HELENA project. Table 1 shows the general workshop schedule. Besides the reports on the current state of the work in the different regions all across the globe, a key activity in the workshop is working in *Break-out Sessions*. These sessions aim at identifying and further developing topics of interest that allow for (i) continuing the survey research, and (ii) to form working groups within the HELENA team. Different to the first workshop, we also provide room for *Lightning Talks* in which HELENA team members and interested "externals" discuss different topics of interest and/or challenge the team and the research findings obtained so far. Finally, this workshop will also continue developing a research agenda to steer further work on the use of hybrid development approaches.

## 4. Conclusion and Future Work

Research conducted in the HELENA community so far strongly indicates the high relevance of the topic. Specifically, the combination of different software and system development approaches has become reality (see Fig. 2) and, moreover, as we could show in [5], it happens to all companies—independent of their size or the respective industry sectors. With this second workshop, we can also add the "region" as a further parameter (Table 1) and, thus, conclude that combination of different development approaches also happens independently from the actual region. That is, hybrid software and system development is a world- wide phenomenon, which requires further attention.

This second workshop is the last one performed during the HELENA stage 2 data collection. The third HELENA workshop will be held in conjunction with the *Evaluation and Assessment in Software Engineering Conference* (EASE) 2018, June 28–29, 2018 in Christchurch, New Zealand.

**Acknowledgements.** We want to thank the Profes 2017 Chairs and organization board for providing us with the opportunity to held the second workshop in conjunction with Profes 2017. We look forward to a fruitful and long-term collaboration with the Profes community.

## REFERENCES

1. Bourque, P., Fairley, R.E. (eds.): Guide to the Software Engineering Body of Knowledge, Version 3.0. IEEE Computer Society, Washington, DC (2014)
2. Brooks, F.P.: No silver bullet essence and accidents of software engineering. IEEE Comput. 20(4), 10–19 (1987)
3. Clarke, P., O'Connor, R.V.: The situational factors that affect the software development process: towards a comprehensive reference framework. Inf. Softw. Technol. 54(5), 433–447 (2012)
4. Kuhrmann, M., Diebold, P., Münch, J., Tell, P., Garousi, V., Felderer, M., Trektere, K., McCaffery, F., Prause, C.R., Hanser, E., Linssen, O.: Hybrid software and system development in practice: waterfall, scrum, and beyond. In: Proceedings of the International Conference on Software System Process, ICSSP, pp. 30–39. ACM, New York, July 2017
5. Kuhrmann, M., Diebold, P., Münch, J., Tell, P., Trektere, K., McCaffery, F., Garousi, V., Felderer, M., Linssen, O., Hanser, E., Prause, C.R.: Hybrid software development approaches in practice: a European perspective. IEEE Softw. (2017, in press)
6. Kuhrmann, M., Münch, J., Diebold, P., Linssen, O., Prause, C.R.: On the use of hybrid development approaches in software and systems development: construction and test of the HELENA survey. In: Proceedings of the Annual Special Interest Group Meeting Projektmanagement und Vorgehensmodelle (PVM). Lecture Notes in Informatics (LNI), vol. P-263, pp. 59–68. Gesellschaft für Informatik (GI), Bonn (2016)
7. Kuhrmann, M., Münch, J., Tell, P., Diebold, P.: Summary of the 1st international workshop on hybrid development approaches in software systems development. ACM SIGSOFT Softw. Eng. Notes (2017, submitted)
8. Theocharis, G., Kuhrmann, M., Münch, J., Diebold, P.: Is Water-Scrum-Fall reality? On the use of agile and traditional development practices. In: Abrahamsson, P., Corral, L., Oivo, M., Russo, B. (eds.) PROFES 2015. LNCS, vol. 9459, pp. 149–166. Springer, Cham (2015). doi:10.1007/978-3-319-26844-6 11
9. West, D.: Water-Scrum-Fall is the reality of agile for most organizations today. Technical report, Forrester (2011)
10. Xu, P., Ramesh, B.: Using process tailoring to manage software development challenges. IT Prof. 10(4), 39–45 (2008)